# Comment on "Improved Analysis of List Decoding and Its Application to Convolutional Codes and Turbo Codes"

by

*Tor Aulin, Fellow IEEE*

Submitted for publication as Correspondence in

*IEEE Transactions on Information Theory*

Suggested Editorial Area: *Coding Theory* (Dr. Robert J. McEliece)

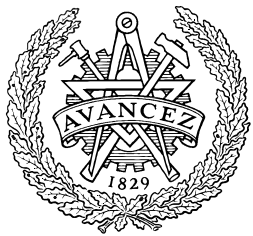

The author is with
Chalmers University of Technology
Telecommunication Theory
SE-412 96 Gothenburg
Sweden
Email: tor@ce.chalmers.se

*Abstract* – In a recent paper [1] an improved analysis concerning the analysis of List Decoding was presented. The event that the correct codeword is excluded from the list is central. For the additive white Gaussian noise (AWGN) channel an important quantity is the in [1] called *effective Euclidean distance*. This was earlier considered in [2] under the name *Vector Euclidean Distance*, where also a simple mathematical expression for this quantity was easily derived for any list size. In [1], a geometrical analysis gives this when the list size is 1, 2 or 3.

The concept of List Decoding (in the Maximum Likelihood sense [6]) is considered in [1] and a central event here is Codeword Error (CE). This means that the correct codeword is not in the list of *L* codewords (using notation in [1]) which are associated with the *L* largest likelihood function values. It is of interest to find the probability of this event for various values of *L*, actually as a function of the signal-to-noise ratio (SNR) $E_b / N_0$. Here $E_b$ is the energy per information bit and $N_0 / 2$ is the double-sided power spectral density of the AWGN.

Using a signal space interpretation [6], one considers the transmitted signal point (vector, codeword) and also *L* other alternatives, codewords. If the received vector is closer to the *L* alternatives than the correct alternative, a CE occurs. Considering the *L* hyper planes half way in between the transmitted signal point and the *L* alternatives, a CE occurs if this received vector is on the far side of all *L* hyper planes, seen from the transmitted signal point. These hyper planes are perpendicular to the vector joining the transmitted signal point and the respective signal alternative vector. The CE region of signal space is the intersection of the half spaces on the remote side of the hyper planes. Integrating the relevant multidimensional probability density function over this region (the CE region) gives the probability of a CE for these *L* alternative vectors.

By considering the minimum Euclidean distance to the CE region, measured from the transmitted signal point, the *effective Euclidean distance* is obtained. This is referred to as the *Vector Euclidean Distance* (VED) in [4],[2]. This quantity will give the asymptotic behavior property of the CE probability (large SNR). This is considered in [1] for the cases *L*=1, 2 and 3. When *L*=3, the special case where the *L*+1 signal points form a tetrahedron is chosen.

When *L*>3, a geometric approach is hard to do (especially to visualize). The approach in [2] uses mathematical statistics and results in a general expression valid for all relevant values of *L* (denoted B in [2]). Further, the number of alternative (incorrect codewords) is in general much larger than *L*, thus all combinations of *L* taken from the total must be considered and the minimum taken, giving the true asymptotic behavior. A general algorithm for accomplishing this is given in [3] for this purpose.

The dimensionality of the signal space spanning the vector from the transmitted signal to the *L* alternatives is at most *L*. It can, however, be lower and can be as low as 1. This, if all vectors are pointing in the same direction. In this case the VED is the largest of the respective distances. All these cases are considered elegantly in [2],[3] by using matrix methods, eigen-system analysis and general Quadratic Programming resulting in simple equations. This method has been used to calculate minimum VEDs for list sizes up to 128 and general trellis codes (where the uniform error property does not hold) [7]. An example is shown in [4] where CPM (Continuous Phase Modulation) systems are considered. It is there of interest to find the minimum size of the list giving the same asymptotic detection performance as if optimal detection (unconstrained list size) was done.

Simulations of specific systems are in [5], showing good agreement with the results using minimum VEDs.

The approach presented in this *Comment* generalizes in almost every aspect the results in [1] (and even more so the references therein). Exact results are obtained for any given situation, not only lower bounds.